\documentclass[12pt]{article}
\usepackage{epsf,latexsym}
\usepackage[all]{xy}
\usepackage{graphicx}
\usepackage{amsfonts,amssymb} 
\epsfverbosetrue
\newcommand{\double}[1]{\mathbb{#1}}

\newcommand{\cc}{\double{C}}

\newcommand{\rr}{\double{R}}
\newcommand{\zz}{\double{Z}}

\newcommand{\aaa}{\mathcal{A}}
\newcommand{\ccc}{\mathcal{C}}

\newcommand{\hhh}{\double{H}}

\newcommand{\hh}{\mathcal{H}}

\newcommand{\llll}{\mathcal{L}}

\newcommand{\ttt}{{\rm tr}}

\def\ddd{{\,\hbox{$\partial\!\!\!/$}}}

\newcommand{\pa}{\partial}

\newcommand{\op}{\oplus}

\newcommand{\bb}{\begin{eqnarray}}
\newcommand{\ee}{\end{eqnarray}}
\newcommand{\eee}{\nonumber\end{eqnarray}}
\newcommand{\qq}{\quad}

\begin{document}

\font\twelve=cmbx10 at 13pt
\font\eightrm=cmr8

\thispagestyle{empty}

\begin{center}
${}$
\vspace{3cm}

{\Large\textbf{The noncommutative standard model, post- and predictions}} \\

\vspace{2cm}

{\large Thomas Sch\"ucker\footnote{also at Universit\'e de Provence, Marseille,
France, thomas.schucker@gmail.com } (CPT\footnote{Centre de Physique
Th\'eorique\\\indent${}$\qq\qq CNRS--Luminy, Case
907\\\indent${}$\qq\qq 13288 Marseille Cedex 9,
France\\\indent${}$\qq
Unit\'e Mixte de Recherche (UMR 6207) du CNRS et des Universit\'es
Aix--Marseille 1 et 2\\
\indent${}$\qq et Sud Toulon--Var, Laboratoire affili\'e \`a la
FRUMAM (FR 2291)})}

\vspace{3cm}

\vskip 2cm

{\large\textbf{Abstract}}
\end{center}
\vskip 1cm
I try to assess the weak and strong points of the standard model of electro-magnetic, weak and strong forces, how it can be derived from general relativity by generalizing Riemannian to noncommutative geometry and what post- and predictions this unification of all four forces entails in particle physics.

\vspace{2cm}\noindent
contribution to Moriond '10 electro-weak
\vskip 1truecm
${}$
\vspace{2cm}

${}$

\section{The standard model}

Let us take a step back and compare atomic and particle physics. The former had started with the discovery that atomic spectra were discrete. Physicists tried to put some order in the fast growing body of experimental numbers. Particularly successful was the ansatz by Balmer and Rydberg for the frequencies $\nu
$ of atomic rays: $\nu=g(n_2^q-n_1^q)$, where $n_1$ and $n_2$ are natural numbers labelling the ray. This ansatz comes with a discrete parameter, the power $q\in\zz$, and with a continuous parameter $g\in\rr$. The experimental fit was successful and determined the two parameters, for instance for the hydrogen atom: $q=-2$ and $g=3.289\ 10^{15}$ Hz, the famous Rydberg constant.  Later the underlying theory was discovered, quantum mechanics. `Underlying theory' means that it allows us to derive the ad hoc ansatz from a first principle, the uncertainty relation. This derivation implies constraints on the parameters, on the nose $q=-2$ and $g=m_e/(4\pi \hbar^3)\ e^4/(4\pi \epsilon_0)^2$.

\subsection{The ansatz}

A new chapter of particle physics had opened with the idea that nuclear forces are mediated by spin 1 particles, the so-called gauge bosons. This idea is formalized using an ansatz that goes by the name of Yang-Mills-Higgs, although other physicists and mathematicians  contributed to its elaboration:
Maxwell, Oskar Klein, Gordon,
 Dirac,  Weyl,
 Elie Cartan, Majorana,
Yukawa, Nambu, Kibble, Goldstone, Brout, Englert,...
The ansatz comes with 4 discrete parameters and a bunch of continuous ones, their number depending on the 4 discrete choices. The first discrete parameter is a real, compact Lie group $G$ and the gauge bosons $A$ live in its complexified Lie algebra. We further need three unitary representations $\rho _\cdot$ on complex Hilbert spaces $\hh_L,\ \hh_R$ and $\hh_S$ to accommodate 
left- and right-handed spinors, $\psi =\psi _L\op\psi _R$ and scalars $\varphi $. Let us recall that since Wigner's classification of the unitary representations of the Poincar\'e group by mass and spin or helicity, particle physicists more or less agree on the general definition of a particle as an orthonormal basis vector in the Hilbert space of a unitary representation of some Lie group. We also need continuous parameters: gauge couplings $g$, scalar self-couplings $\lambda $ and $\mu $, and complex Yukawa couplings $g_Y$. In some cases we may add gauge-invariant (Majorana-) masses. Once we have chosen all these parameters, we can start business and write down the ansatz, the Yang-Mills-Higgs Lagrangian.
\begin{center}
\begin{tabular}{llr}
$\
\llll[A,\psi ,\varphi ]=$&${}\ {\textstyle\frac{1}{2}}\, \ttt(\pa_\mu
A_\nu
\pa^\mu A^\nu -
\pa_\mu A_\nu \pa^\nu A^\mu)$& 
\includegraphics[viewport=8.8cm 14.2cm 11cm 16cm,width=1.5cm, height=1cm]{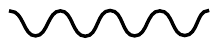}
\\
&$+g\ \ttt (\pa_\mu A_\nu [A^\mu ,A^\nu ])$&
\includegraphics[viewport=8.5cm 13.4cm 12.5cm 17cm,width=1.5cm, height=0.8cm]{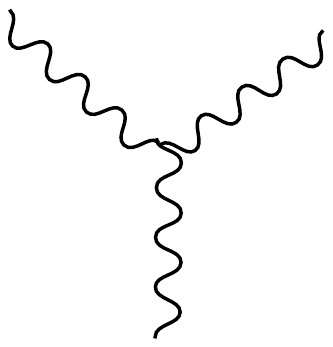}
\\
&$+g^2\,\ttt ([A_\mu ,A_\nu ][A^\mu ,A^\nu ])$&
\includegraphics[viewport=8.5cm 13.6cm 12.5cm 17cm,width=1.5cm, height=0.8cm]{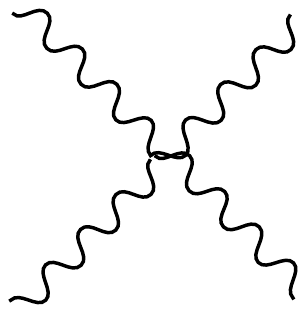}
\\[-4mm]
&$ +\bar \psi \ddd\psi  $&
\includegraphics[viewport=8.8cm 15.3cm 11cm 17.2cm,width=1.5cm, height=1cm]{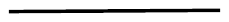}
\\[-2mm]
&$ +i g\,\bar\psi (\tilde\rho _L\op\tilde\rho _R)(A_\mu )\,\gamma
^\mu \psi  $&
\includegraphics[viewport=8.5cm 13.5cm 12.5cm 17cm,width=1.5cm, height=0.8cm]{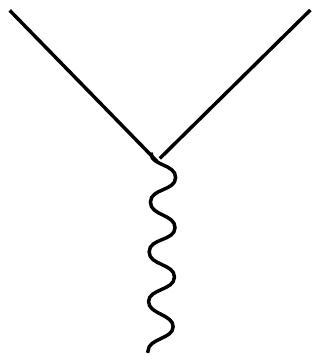}
\\[-4mm]
&
$+ {\textstyle\frac{1}{2}} \,\pa_\mu \varphi ^*\pa^\mu\varphi  $
&
\includegraphics[viewport=7.2cm 14.9cm 9.5cm 16cm,width=1.5cm, height=1cm]{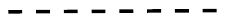}
\\&
$+ {\textstyle\frac{1}{2}} \,g\,\{(\tilde \rho _S(A_\mu )\varphi)
^*\pa^\mu \varphi  + \pa_\mu \varphi ^*
\tilde \rho _S(A_\mu )\varphi \}$\hspace{4mm} &
\includegraphics[viewport=8.5cm 13.5cm 12.5cm 17cm,width=1.5cm, height=0.8cm]{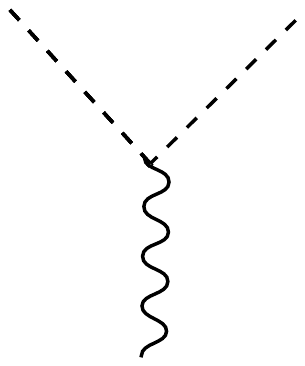}
\\
&$ + {\textstyle\frac{1}{2}} \,g^2\,(\tilde \rho _S(A_\mu )\varphi)
^*\tilde \rho _S(A^\mu )\varphi$
&
\includegraphics[viewport=4.8cm 17.6cm 8.7cm 21cm,width=1.5cm, height=0.8cm]{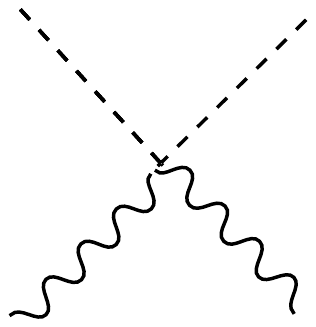}
\\
&$ +\lambda \, \varphi ^*\varphi \varphi ^*\varphi
$&
\includegraphics[viewport=8.5cm 13.6cm 12.5cm 17cm,width=1.5cm, height=0.8cm]{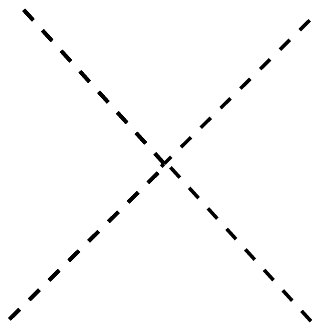}
\\[-8mm]
&$ -{\textstyle\frac{1}{2}} \,\mu ^2\,\varphi ^*\varphi
$&
\includegraphics[viewport=9.2cm 15cm 11.7cm 16cm,width=1.5cm, height=1.3cm]{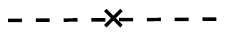}
\\[-2mm]
 &$ +\,g_Y\,\bar\psi \varphi \psi
+\,\bar g_Y\,\bar\psi \varphi ^* \psi $&
\includegraphics[viewport=8.5cm 13.5cm 12.5cm 17cm,width=1.5cm, height=0.8cm]{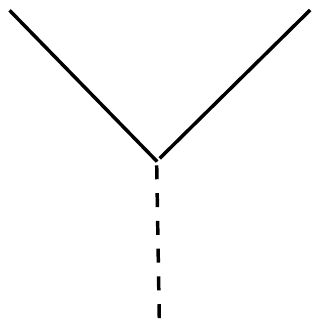}
\end{tabular}
\end{center}\vspace{4mm}
This Lagrangian has four important properties:\\[2mm]
$\bullet$ 
If we choose the group abelian, $G=U(1)$, avoid parity violation, $\hh_L=\hh_R$, and suppress\\ ${}\ \, $
 scalars, $\hh_S=\{0\}$, we retrieve Maxwell's Lagrangian 
with  $g=e/\sqrt{\epsilon_0}$.\\
$\bullet$
The Yang-Mills-Higgs Lagrangian is gauge invariant for the group $G$.\\
$\bullet$
Gauge invariance in turn implies that the Lagrangian defines a perturbatively renormalizable\\ ${}\ \, $
 quantum field theory on flat Minkowski space, if in addition the Yang-Mills anomaly vanishes.\\
$\bullet$
As a classical field theory, the Yang-Mills-Higgs action properly coupled to general relativity \\ ${}\ \, $
 is invariant under general coordinate transformations.

\subsection{The fit}

The experimental fit of the Yang-Mills-Higgs Lagrangian to an impressive body of millions of experimental numbers (cross-sections, branching ratios, life-times, ...) is successful and yields the following discrete parameters:
\bb G&=&SU(2)\times U(1)\times
SU(3)/(\zz_2\times\zz_3).\ee
Note the quotient by a discrete subgroup in the center, first observed by O'Raifeartaigh \cite{or}. 
The particle content of $G$ is
\bb {\rm Lie}(G)^\cc&\owns&\left( W^\pm,\, Z,\,{\rm photon}\right) \oplus \left( \,8\ {\rm gluons}\right).\ee
The left-handed spinors form a highly reducible representation,
\bb \hh_L &=& \bigoplus_1^3\left[ 
(2,{\textstyle\frac{1}{6}},3)\op
(2,-{\textstyle\frac{1}{2}},1)
\right]   \label{HL}
 \\[1mm]
&\owns& \left[ 
\left( \begin{array}{c}
u\\ d
\end{array}\right)_L\oplus 
\left( \begin{array}{c}
\nu_e\\ e
\end{array}\right)_L\right] \oplus
\left[ 
\left( \begin{array}{c}
c\\ s
\end{array}\right)_L\oplus 
\left( \begin{array}{c}
\nu_\mu \\ \mu 
\end{array}\right)_L\right] \oplus
\left[ 
\left( \begin{array}{c}
t\\ b
\end{array}\right)_L\oplus 
\left( \begin{array}{c}
\nu_\tau\\ \tau
\end{array}\right)_L\right].\nonumber\ee
We denote by $(2,{\textstyle\frac{1}{6}},3)$ the tensor product of a doublet under $SU(2)$ and a triplet under $SU(3)$ with $U(1)$ hypercharge ${\textstyle\frac{1}{6}}$.
The right-handed Hilbert space is even more reducible,
\bb\hh_R& = &\bigoplus_1^3\left[ 
(1,{\textstyle\frac{2}{3}},3)\oplus
(1,-{\textstyle\frac{1}{3}},3)\op (1,-1,1)\op (1,0,1)
\right] \label{HR}
 \\[1mm]
&\owns& \left[ 
u_R\oplus d_R\oplus e_R
\oplus 
\nu_{eR}
\right] \oplus\left[ 
c_R\oplus s_R\oplus \mu _R
\oplus 
\nu_{\mu R}
\right] \oplus\left[ 
t_R\oplus b_R\oplus \tau_R
\oplus 
\nu_{\tau R}
\right] .\nonumber\ee
It is a miracle that the fermionic representation $\hh :=\hh_L\op\hh_R$ has vanishing Yang-Mills and also vanishing gravitational anomalies. Finally the scalar representation is choosen irreducible, an isospin doublet, colour singlet, 
\bb\hh_S &= &(2,-{\textstyle\frac{1}{2}},1)\qq
 \owns\ \left( W^\pm_{\rm long},\, Z_{\rm long},\,{\rm Higgs}\right).\ee
This choice follows the wish to render massive only the three weak gauge bosons $W^\pm,\, Z$ and implies the existence of only one physical scalar, the `Higgs' $H$. 

For the continuous parameters, the experimental fit gives $g_2=0.6518\pm 0.0003,$ $g_1=0.3574\pm 0.0001$ and  $g_3=1.218\pm 0.01$ for the gauge couplings at $E=m_Z$. Concerning the scalar self-couplings $\lambda $ and $\mu$, we have today only  one value and one bound:
\bb m_W=\,\frac{g_2\mu }{4\sqrt{\lambda }}\, =80.398\pm 0.025\ {\rm GeV},&m_H=\sqrt{2}\mu >114.4\ {\rm GeV}.\ee
Finally, the Yukawa couplings can be traded for the Dirac masses and mixings and we may also add Majorana masses and mixings for the right-handed neutrinos because they are neutral under the entire group $G$.
\subsection{The theory}
To complete our comparison of atomic and particle physics, we need the underlying theory, which is noncommutative geometry \cite{gnc}. Indeed it derives the complicated Yang-Mills-Higgs ansatz from first principles: geometry and general relativity. Again, this derivation implies constraints of which the most spectacular certainly is: the scalar representation is computed not chosen. And for the standard model, this computation produces, on the nose, the scalar representation chosen by experiment, $\hh_S = (2,-{\textstyle\frac{1}{2}},1)$.
\vspace{2mm}
\begin{center}
\begin{tabular}{l|ll}
& atomic physics& 
particle physics\\[1ex]
\hline &\\
new physics& discrete spectra&
forces from gauge
bosons\\[1ex]
ansatz &  $\nu ={g}\,({n_2}^{q}-{n_1}^{q})$& 
Yang-Mills-Higgs
models
\\
&Balmer-Rydberg&
 \\[1ex]
discrete param.&
${ q}\in \zz$ &
{$G,\,\hh_L,\,\hh_R,\,\hh_S$}
\\[1ex] 
continuous param.&
${g}\in \rr_+$ &
${g,\,\lambda,\,\mu}\in \rr_+,\ { g_Y}\in\cc $
\\[1ex] 
experimental fit& ${q}=-2,$
&standard model\\
&${g}=3.289\ 10^{15}$ Hz&
\\[1ex] 
underlying theory  &  quantum
mechanics&
noncommutative geometry
\\[1ex] 
constraints &${q}=-2$ & 
${\hh_S}=(2,-{\textstyle\frac{1}{2}} ,1)$, ...
\\[1ex] 
&${g}=\,\frac{m_e}{4\pi \hbar ^3}\,\frac{e^4}{(4\pi \epsilon_0)^2}\, $& 
${g_2}^2={g_3}^2=3{\lambda}={\textstyle\frac{1}{4}}\sum |{g_Y}|^2  $
\end{tabular}
\end{center}
\vspace{4mm}

\section{The noncommutative version of the standard model}

\subsection{Noncommutative geometry}

It is easy to give an avant-go\^ut of 
noncommutative geometry to physicists. Indeed, historically the first
 example of a noncommutative geometry is quantum mechanics. It equips classical phase-space $\owns\, (x,p)$ with Heisenberg's uncertainty relation $\Delta x\Delta p \ge {\textstyle\frac{1}{2}} \hbar$ by rendering the commutative algebra of classical observables $\aaa=\ccc^\infty ($phase space) noncommutative. A classical observable is a differentiable function of $x$ and $p$ like the energy of the harmonic oscillator $p^2/(2m)+kx^2/2$. By putting $ [x,p]=i\hbar 1$ this classical observable is promoted to an operator, i.e. an element of the noncommutative algebra of quantum observables $\aaa$. Both algebras are associative. In the quantum case, $\aaa$ must have an involution (to define self-adjoint operators with real spectrum) and a faithful, unitary representation on a Hilbert space $\hh$ whose elements are the wave-functions $\psi $. Furthermore the dynamics in relativistic quantum mechanics is defined by the Dirac operator $\ddd$, a self-adjoint operator on $\hh$. ($\ddd$ is not in the algebra.) Quantum mechanics is defined by the three building blocks $(\aaa,\hh,\ddd)$ and so is a general noncommutative geometry \cite{gnc}. Connes calls them a `spectral triple'. The general definition makes sense thanks to Connes' reconstruction theorem \cite{recon} that he published on the hep-th arXiv in 1996. The theorem establishes a one-to-one correspondence beween {\it commutative} spectral triples (i.e. $\aaa$ commutative) and Riemannian spin manifolds. It therefore offers a reformulation of Riemannian geometry in the operator algebraic language of quantum mechanics. Like in quantum mechanics, a noncommutative space is defined by a noncommutative spectral triple. Intuitively noncommutative geometry does to curved space(time) what quantum mechanics did to (flat) phase space, equipping space with an uncertainty relation.
 
 Then Connes remarks that noncommutative spaces are close enough to Riemannian spaces such that Einstein's derivation of gravity from Riemannian geometry carries over to noncommutative spaces. In Connes' derivation, the entire Yang-Mills-Higgs action pops up as a companion to the Einstein-Hilbert action, just like the magnetic field pops up as a companion to the electric field, when the latter is generalized to Minkowskian geometry, i.e. special relativity. And Connes derivation implies constraints. On the discrete side, besides being able to compute the scalar representation $\hh_S$ we get constraints on the choice of the group $G$ and of the fermionic representation $\hh=\hh_L\oplus\hh_R$.
 
 \subsection{Constraints on discrete parameters}
 
 In noncommutative geometry, the invariance group is the automorphism group of the algebra $G={\rm Aut}(\aaa)$ lifted to the Hilbert space. In the commutative case, the reconstruction theorem tells us that the algebra is the algebra $\ccc^\infty(M)$ of differentiable functions on the space(time) $M$, just as the algebra of classical observables. Its automorphism group is on the nose the group of general coordinate transformations, 
 ${\rm Aut}(\ccc^\infty(M))={\rm Diff}(M)$, that characterize general relativity. To go noncommutative, let us take the cheapest noncommutative algebra on the market, the quaternions $\hhh$. Physicists know this algebra well from spin 1/2 and they also know that ${\rm Aut}(\hhh)=SU(2)/\zz_2$. Another example is the algebra of complex $3\times 3$ matrices whose automorphism group is $SU(3)/\zz_3$. Counter-examples are all the exceptional groups and $U(1)$, which are not automorphism groups of associative algebras. However, $U(1)$ re-surfaces through the spin-lift to $\hh$. 
 
 The constraints on the fermionic representation come from the fact that the group representation from Yang-Mills must extend to an algebra representation. This leaves us essentially with fundamental and singlet representations and direct sums thereof, `generations'.  This constraint is indeed fulfilled by the standard model, see equations (\ref{HL}) and (\ref{HR}). Then another constraint from geometry tells us that -- restricted to $SU(3)$ -- $\hh_L$ and $\hh_R$ must coincide. This is again true in equations (\ref{HL}) and (\ref{HR}), but in Connes' approach it is not opportunistically arranged by hand in order to preserve parity under the strong force and to keep experimentalists satisfied, it is dictated by geometry.
 
 \subsection{Constraints on continuous parameters}
 
 Now that the standard model has passed all tests from the discrete parameters, let us turn to the constraints on its continuous parameters. They are given by \cite{cc}:
 \bb {g_2}^2={g_3}^2=3{\lambda}={\textstyle\frac{1}{4}}\sum |{g_Y}|^2. \ee 
 We can make sense out of these equations if we assume the ill-famed hypothesis of the big desert and if we assume the perturbative renormalization group flow on flat and commutative Minkowski space with $\Lambda =10^{17}$ GeV, see figure.
 \vspace{4mm}
 \begin{center}
\includegraphics[width=11cm, height=5cm]
{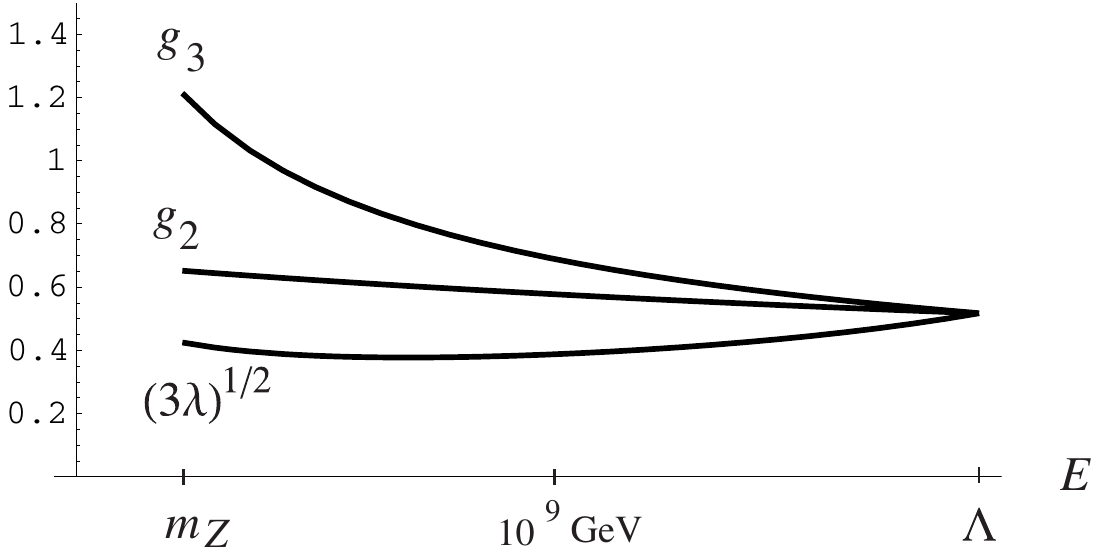}
\end{center}
 \vspace{2mm}
To conclude, let us list the post- and predictions that follow from Connes' geometry:
\\[2mm]
$\bullet$ {\it Prediction:\ }  Fermions cannot live in weak isospin triplet or higher representations. \\
$\bullet$ {\it Prediction:\ }
There is only one Higgs scalar. (Supersymmetry has 5.) \\
$\bullet$ {\it Postdiction:}
The  strong force preserves parity.\\
$\bullet$ {\it Postdiction:}
Photons and gluons are massless.\\
$\bullet$ {\it Postdiction:}
The $\rho_0$ parameter
$  :=(g_1^2+g_2^2)/g_1^2\  m_W^2/m_Z^2\, =1\ \left( \ =1.0002^{+0.0007}_{-0.0004},\ {\rm exp.}\right). $\\
$\bullet$ {\it Prediction:\ }
$\,g_2=g_3$ plus big desert implies uncertainty in proper-time measurements\\ ${}\qq\qq\qq\qq\qq\qq\ \, \Delta \tau\sim\hbar/\Lambda \sim 10^{-41}$ s.\\
$\bullet$ {\it Prediction:\ }
$\,g_2=\sqrt{3\lambda }$ plus big desert implies
$ m_H=170\pm10\ {\rm GeV}.$\\
$\bullet$ {\it Postdiction:}
$\,g_2={\textstyle\frac{1}{2}} (\sum |g_Y|^2)^{1/2}$ plus big desert implies
$ m_t<186\ {\rm GeV}$.\\ ${}\qq\qq\qq\qq\qq\qq\ $ ($ m_t=171.3\pm2.3\ {\rm GeV}$,\ {\rm exp.})\\
$\bullet$ {\it Prediction:\ }
 $\,g_2={\textstyle\frac{1}{2}} (\sum |g_Y|^2)^{1/2}$  excludes a fourth generation.

\section{Loose ends}

\subsection{Quantum fields}

Quantum mechanics is a noncommutative geometry, the so-called Moyal plane \cite{qm}. The standard model is a noncommutative geometry and it unifies naturally with the commutative geometry of general relativity. However it must be emphasised that this unification is non-quantum, i.e. it occurs at the level of classical field theory. A major motivation for a noncommutative geometry on spacetime is to cure the short-distance divergencies of quantum field theory and already Heisenberg had this dream. Despite two technical break-throughs, a conceptual formulation of minimal subtraction in the language of noncommutative geometry \cite{ck} and a renormalizable scalar field theory on the Moyal space \cite{gw}, the dream of quantum field theory on curved and noncommutative spaces is still out of reach. Meanwhile we copy from our colleagues. They say: ``It is true that we do not have a coherent theory of quantum fields in curved space yet. However general relativity indicates that curvature becomes relevant only at energies above the Planck mass, $10^{19}$ GeV. We already have a pretty hard time to reach $10^{3}$ GeV and we will therefore stick to quantum field theory in flat Minkowski space, a theory on which a comfortable majority seems to agree and from which a more or less coherent picture emerges.'' So, we say: ``It is true that we do not have a coherent theory of quantum fields in noncommutative space yet. However the standard model indicates that noncommutativity becomes relevant only at energies above $\Lambda =10^{17}$ GeV. ...''

\subsection{Noncommutative geometry beyond the standard model}

The standard model fits naturally into the frame of noncommutative geometry, the minimal spectral triplet being the standard model with one generation of quarks and leptons without a right-handed neutrino. On the other hand it is extremely difficult to go beyond the standard model in this frame. Christoph Stephan put the classification of finite spectral triplets on the computer and found two examples beyond the standard model that are not in conflict with experimental evidence: The first one \cite{cs} has an additional, unbroken $SU(4)$ gauge group, confined $SU(4)$-singlets in the TeV range, gauge unification at $\Lambda =10^{4}$ GeV  and a Higgs-mass of $241\pm2$ GeV. The second one \cite{ss} adds three vectorlike iso-spin doublets to the standard model. It has gauge unification at $\Lambda =10^{7}$ GeV  and a Higgs-mass of $203\pm2$ GeV.

\subsection{Other Higgs-mass predictions}

In the literature you can find today over a hundred predictions for the Higgs-mass testifying to theorists' imaginativeness and obstinacy. By far the most precise prediction comes from E-infinity theory \cite{el} and has a Higgs of 161.8033989  GeV. Forty predictions alone come from mildly extended minimal supersymmetric models. As a service to our Higgs hunting community, I maintain a compilation of these predictions \cite{comp}.

\end{document}